\documentclass{article}

\usepackage[utf8]{inputenc}
\usepackage[english]{babel}
\usepackage{amssymb}
\usepackage{amsmath}
\usepackage{graphicx}
\usepackage{subfigure}
\usepackage{makeidx}
\usepackage{multicol}
\usepackage{inputenc}
\usepackage{epigrafica}
\usepackage[LGR,OT1]{fontenc}
\usepackage{mathptmx}
\usepackage{multirow}
\usepackage{hhline}
\usepackage{abstract}
\usepackage{amsmath}
\usepackage{graphicx}

\title{A comparative study of self-starting CUSUM control charts for location shifts}
\author{Konstantinos Bourazas \\
Department of Economics, \\
Athens University of Economics and Business, Athens, Greece \\
        \texttt{kbourazas@aueb.gr}}
\date{} 

\begin{document}

\maketitle

\begin{abstract}
In recent years, self-starting methods have garnered increasing attention in Statistical Process Control and Monitoring (SPC/M), as they offer real-time disorder detection without the need for a calibration phase (Phase I). This study focuses on evaluating parametric self-starting CUSUM-type control charts, specifically the Bayesian Predictive Ratio CUSUM (PRC) developed by Bourazas et al. (2023) and the frequentist alternative self-starting CUSUM proposed by Hawkins and Olwell (1998). The performance of these methods is thoroughly examined through an extensive simulation study under various scenarios involving a change in the mean of Normal data. Additionally, a prior sensitivity analysis for PRC is conducted. The work ands with concluding remarks summarizing the findings. 

\textbf{keywords}: CUSUM, location shift, PRC, self-starting, SSC, statistical process control and monitoring
\end{abstract}

\section{Introduction}

Statistical Process Control and Monitoring (SPC/M) is an area of Statistics whose primary objective is the real-time detection of anomalies in an ongoing process. Specifically, the goal is to identify when a process transitions from its In Control (IC\index{IC}) state, where it operates only under natural causes of variation, to the Out Of Control (OOC\index{OOC}) state, where assignable causes of variation are present \cite{montgomery2019introduction}. The implementation of SPC/M methods is mainly performed via control charts, while most of these methods involve two phases. {Phase I}\index{Phase I}, which is the training phase and involves the collection of independent IC data to calibrate the monitoring scheme, aiming to derive reliable estimates of the distribution and unknown parameters. Then, in {Phase II}\index{Phase II}, i.e. the testing phase, we collect new observations for comparison against the IC standards established in Phase I. The limitations of the Phase I/II separation are well-documented, mainly tied to assumptions and procedures during Phase I.  {For instance, the estimation error of Phase I estimates, which negatively affects performance during Phase II, is typically not taken into account (see \cite{qiu2013introduction}). To eliminate this impact, the Phase II data could be used to update the Phase I estimates. However, this available information is usually wasted, as the learning stops at the end of Phase I.} Additionally, if a shift in a parameter occurs during Phase I, it can deteriorate the performance of both phases. An extensive discussion of Phase I methods is provided in \cite{jones2014overview}, highlighting major issues and advancements. Self-starting methods in SPC/M aim to mitigate issues related to Phase I/II separation. These methods do not require a calibration phase (Phase I) to estimate the unknown parameters, but they can provide testing from the start of a process while estimating the unknown parameters. 

Self-starting control charts have been developed for both of the two major parametric approaches in Statistics, the frequentist\index{frequentist} and the Bayesian\index{Bayesian}. The concept of self-starting control charts was introduced in \cite{hawkins1987self}, where a frequentist CUSUM-type control chart for location and scale shifts was developed. Staying within the frequentist framework, notable are the self-starting CUSUMs (SSCs, \cite{hawkins1998cumulative}) that are widely used. Additional noteworthy frequentist self-starting methods for monitoring the process mean are presented in works such as \cite{hawkins2003changepoint}, \cite{hawkins2005statistical}, or \cite{tercero2014self}. From the Bayesian perspective, a family of Bayesian self-starting control charts based on the posterior predictive distribution introduced in \cite{bourazas2023predictive}, while Bayesian self-starting methods for monitoring the mean of a process can be found in  \cite{tsiamyrtzis2005bayesian}, \cite{tsiamyrtzis2008bayesian}, and \cite{tsiamyrtzis2010bayesian}. Furthermore, we should highlight the online change point in \cite{adams2007bayesian}, who employed a machine learning algorithm to estimate a change point, utilizing the posterior distribution of the length of an IC sequence. Concerning the outlier detection, numerous self-starting methods have been proposed, with the frequentist $Q$ charts in \cite{quesenberry1991aspc}
, \cite{quesenberry1991bspc}, and \cite{quesenberry1991cspc} for Normal, Binomial, and Poisson data, and the Bayesian Predictive Control Charts \cite{bourazas2022predictive} to be representative. 

In this work, we focus on the comparison of two well-established self-starting parametric methods for univariate data, the frequentist Self-Starting CUSUM \cite{hawkins1998cumulative} and the Bayesian Predictive Ratio CUSUM \cite{bourazas2023predictive} for persistent location shifts in Normal data. In Section \ref{sec:2}, we briefly present the assumptions and the testing statistics of the methods. Section \ref{sec:3} follows with an extensive simulation study, while Section \ref{sec:4} concludes this work.

\section{Competing methods} \label{sec:2}

In this section, we will provide the assumptions, and the testing procedures of the two competing methods; the frequentist self-starting CUSUM (SSC, \cite{hawkins1998cumulative}) and the Bayesian Predictive Ratio CUSUM (PRC, \cite{bourazas2023predictive}). The under-study control charts are focused on detecting shifts in the mean of Normal data, assuming both the mean and the variance of the process are unknown.

\subsection{Self Starting CUSUM (SSC)} \label{subsec:21}

We start presenting competing methods with SSC\index{SSC}. Assume $\mathbf{X}_n=(x_1,...,x_n)$ to be a random sample from a Normal distribution with unknown mean and variance. Let $\bar{x}_n = \sum_{j=1}^{n} x_j$ and $s_n = \sum_{j=1}^{n} (x_j - \bar{x}_n)^2 / (n-1)$ be the sample mean and the sample variance of $\mathbf{X}_n$ respectively. For the next observable $X_{n+1}$, we define its standardized form as
\begin{align} \label{eq:Tn}
	T_{n+1} = \dfrac{X_{n+1}-\bar{x}_n}{s_n}.
\end{align}
It is known that $\sqrt{\dfrac{n}{n+1}}T_{n+1}\sim t_{n-1}$, i.e. a Student-$t$ distribution with $n-1$ degrees of freedom. Finally, we transform the $T_{n+1}$ into a random variable $U_{n+1}$ that follows an exact standard Normal distribution and it is 
\begin{align} \label{eq:Un}
	U_{n+1} = \Phi^{-1}F_{n-1} \left( \sqrt{\dfrac{n}{n+1}} T_{n+1} \right), 
\end{align}
where $\Phi^{-1}(\cdot)$ is the inverse cumulative function of a standard Normal distribution and $F_{n-1}$ is the cumulative function of a Student-$t$ distribution with $n-1$ degrees of freedom. The CUSUM statistic for $x_{n+1}$ will be 
\begin{align}\label{eq:cumssc}
	C_{n+1}^+& =\max\{0,C_n+U_{n+1}- {k_{SSC}} \}, \;\; or \nonumber \\
	C_{n+1}^-& =\min\{0,S_n+U_{n+1}+ {k_{SSC}} \},
\end{align}
when we are interested in detecting upward or downward shifts respectively and setting the CUSUM statistics for $n=1$ or $2$ equal to zero. The design parameter  {$k_{SSC}$} is associated with the magnitude of the shift targeted for detection, and it is conventionally set to half the size of the shift we want to detect.  {We ring an alarm for an upward or downward shift if $C_{n+1}^+>h_{SSC}^+$ or $C_{n+1}^+<h_{SSC}^-$, respectively. For a two-sided SSC (and assuming $h_{SSC}^+=-h_{SSC}^-=h_{SSC}$) we can plot $C_{n+1} = \max \{ C_{n+1}^+, |C_{n+1}^-| \}$ and an alarm is ringed whenever the statistic $C_{n+1}$ exceeds $h_{SSC}$.}

\subsection{Predictive Ratio CUSUM (PRC)} \label{subsec:22}

Now, we continue with PRC\index{PRC}, which appears to be the Bayesian analog of SSC. PRC is based on the ratio of two posterior predictive distributions, one that results directly from the data and represents the IC state, and a shifted version of the IC predictive that represents the OOC state. Let $X_i|\left( \theta_1, \theta_2^2 \right) \sim N \left(\theta_1,\theta_2^2 \right)$, with $i=1,...,n$ and both parameters being unknown.  {As the joint prior distribution of the mean and the variance, we assume a Normal Inverse Gamma, and specifically $\left( \theta_1, \theta_2^2 \right) \sim NIG \left(\mu_0,\lambda, a,b \right)$. The choice of NIG is convenient as it is conjugate to the Normal likelihood leading to closed-form posterior predictive distributions. NIG can be considered as a hierarchical prior set-up in the mean $\theta_1$ and the variance $\theta_2^2$, where $\theta_1 | \theta_2^2 \sim N(\mu_0, \theta_2^2 / \lambda)$ and $\theta_2^2 \sim IG (a, b)$. The hyperparameters $\mu_0$, $\lambda$,  $a$, and $b$ are considered to be known with fixed values. Regarding the prior hyperparameters determination, a standard way to set is by the desired values for the mean and variance of the priors. However, the prior hyperparameters can also have further interpretation through imaginary data use. In that case, we could estimate $\theta_1$ from $\lambda$ observations with a mean $\mu_0$ and $\theta_2^2$ from $2\alpha$ observations with a sum of the squared deviations of $2\beta$. These relationships arise directly from the form of the posterior of the imaginary data, which will be again a NIG due to the conjugacy. We present this rationale for deriving the hyperparameters of the weakly informative prior used in Section \ref{sec:3}. For more information on the interpretation of prior parameters, see \cite{gelman2013bayesian}. Additionally, the power priors by \cite{ibrahim2000power} are advantageous and flexible as they allow the direct use of historical data from similar processes and their combination with other sources of information. Some guidelines and suggestions regarding the determination of the prior hyperparameters for control charts are provided by \cite{bourazas2022predictive}. In case of prior ignorance, the non-informative reference prior \cite{berger2009formal} utilization is suggested. The reference prior maximizes the expected Kullback–Leibler divergence of the posterior distribution relative to the prior. For this model the reference prior is $\pi \left( \theta_1, \theta_2^2 \right)\propto 1/{\theta_2^2} \equiv NIG(0,0,-1/2,0)$. It is worth mentioning that, even if this prior is improper (it does not integrate to 1), it leads to a proper posterior.}

If we denote $\mathbf{X}_n=(x_1,...,x_n)$, then the IC posterior and predictive distributions will be $\left( \theta_1, \theta_2^2 \right) | \mathbf{X}_{n} \sim NIG \left(\hat{\mu}_{n}, \hat{\lambda}_{n}, \hat{a}_n,\hat{b}_n  \right) $ and $ f \left( X_{n+1} | \mathbf{X}_n \right)=t_{2\hat{a}_{n}}\left( \hat{\mu}_{n},  \left( \lambda_{n} +1 \right) \cdot \hat{b}_{n} \big/ \left( \lambda_{n} \cdot \hat{a}_{n} \right) \right)$ respectively, where $t$ denotes the Student-$t$ distribution and 
\begin{align} \label{eq:postic}
	\hat{\mu}_{n} = \left( \lambda \mu_0 + \sum\limits_{j=1}^{n} x_j \right) \bigg/ \left( \lambda + n   \right),\hat{\lambda}_{n} =  \lambda + n, \hat{a}_{n} = a + n \big/2 \nonumber \\
	\hat{b}_{n} =  b + \left( \lambda \mu_0^2 + \sum\limits_{j=1}^{n} d_j^2 \right) \bigg/ 2  -  \left(  \lambda \mu_0 + \sum\limits_{j=1}^{n} d_j \right) ^2 \bigg/ \left( 2 \left( \lambda + n  \right)  \right).   
\end{align}
To express the OOC shift we introduce a step change of size $ {k_{PRC}} \cdot \hat{\theta}_2$ to the location parameter of the posterior marginal distribution of the mean, where $\hat{\theta}_2 = \sqrt{\hat{b}_{n} / \hat{a}_{n}} $. The shift will be upward or downward if $ {k_{PRC}}>0$ or $ {k_{PRC}}<0$ respectively. Thus, the location parameter of the OOC (intervened) posterior $\hat{\mu}_{n} +  {k_{PRC}} \cdot \hat{\theta}_2$ and the corresponding OOC predictive will be $f ' \left( X_{n+1} | \mathbf{X}_n \right)=t_{2\hat{a}_{n}}\left( \hat{\mu}_{n} +  {k_{PRC}} \cdot \hat{\theta}_2, \left( \hat{\lambda}_{n} +1 \right) \cdot \hat{b}_{n} / \left( \hat{\lambda}_{n} \cdot \hat{a}_{n} \right) \right)$. The logarithmic predictive ratio $log \left( L_{n+1} \right) $ for the next observable $x_{n+1}$ will be 
\begin{align} \label{eq:logprc}
	log \left( L_{n+1} \right) & = \dfrac{f ' \left( X_{n+1} | \mathbf{X}_n \right)}{f  \left( X_{n+1} | \mathbf{X}_n \right)} = \left( \hat{a}_{n} + 1/2 \right) \cdot log  \dfrac{ 2 \hat{a}_{n} + Z_{n+1}^2}{ 2\hat{a}_{n}   + \left(Z_{n+1} -  {k_{PRC}} \cdot \hat{\lambda}_n  / (\hat{\lambda}_{n}+1) \right)^2}, 
\end{align}
where $Z_{n+1} = \left( X_{n+1} - \hat{\mu}_n \right) / \sqrt{ \left( \hat{\lambda}_{n} +1 \right) \cdot \hat{b}_{n} / \left( \hat{\lambda}_{n} \cdot \hat{a}_{n} \right)} $. Note that the design parameter  {$k_{PRC}$} has a different interpretation from  {$k_{SSC}$}; we typically set it to be equal to the size of the shift we want to detect, e.g., for a change of 1 standard deviation we set  {$k_{PRC}=1$}. The CUSUM statistic for the next data point at time $n+1$ will be 
\begin{align}\label{eq:cumprc}
	S_{n+1}^+& =\max\{0,S_n+log \left( L_{n+1} \right) \}, \;\; or \nonumber \\
	S_{n+1}^-& =\min\{0,S_n-log \left( L_{n+1} \right) \},
\end{align}
when we are interested in detecting upward or downward shifts respectively, and setting the first CUSUM statistics equal to 0, or the first two CUSUM statistics when using the reference prior.  {Like SSC, a alarm is raised when $C_{n+1}^+>h_{PRC}^+$ or $C_{n+1}^+<h_{PRC}^-$, upwards and downwards respectively. For two-sided procedures (and assuming $h_{PRC}^+=-h_{PRC}^-=h_{PRC}$), we raise an alarm if $S_{n+1} = \max \{ S_{n+1}^+, |S_{n+1}^-| \} >h$.}


\section{Simulations} \label{sec:3}

In this section, we will evaluate and compare the performance of the methods described in Section \ref{sec:2}; namely the PRC and SSC. The two competing methods had been compared in \cite{bourazas2023predictive} for short runs. However, in this study, we will evaluate them for long runs using the appropriate  {performance} metric.  {Starting from the false alarm tolerance, we configure the control charts for two-sided shifts, and we set the decision limits $h_{SSC}$ and $h_{PRC}$ to achieve an IC Average Run Length ({$ARL_0$}\index{$ARL_0$}) of $ARL_0=370$ for both methods. Algorithmically, we determine the decision limits using the Regula Falsi method for 10,000 simulated sequences from the standard Normal, similar to the approach described in Algorithm 2 in \cite{bourazas2023design}.} For the IC state, we assume that the data follow a standard Normal distribution, while for the OOC sequences, the post-change data come again from a Normal distribution with a shift in the mean of size $\delta$. To assess the performance of the methods for different magnitudes of jumps, we consider four values for the step change: $\delta=0.5$, 1, 1.5, or 2. Continuing with the location of change point $\tau$, we assume different scenarios, where $\tau$ appears from very early to later stages in the data sequence, examining in this way its impact to the detection  {perfomance}. Specifically, we set 10 values for $\tau$ with $\tau=11$, 21, ..., 101. This translates that for the smallest value of $\tau$, the change point occurs after only 10 IC data points available, while for the last value of $\tau$, we have an IC history of 100 data points. As OOC metric, we employ the Conditional Expected Delay (CED\index{CED}), which is given by the formula:

\begin{align} \label{eq:CED}
	CED(\tau) = E_\tau \left( T-\tau+1 | T \geq \tau \right) = \dfrac{E\left( T-\tau+1\right)^+}{P(T \geq \tau)},        
\end{align}
 {where $T$ is the stopping time and $x^+ = \max(0,x)$. $T$ is defined for each method as the first time that its cumulative statistic plots beyond the corresponding decision threshold, and the $CED(\tau)$ is the average delay of an alarm given that the change point has occurred. For every OOC scenario, the $CED(\tau)$ estimate is based on 10,000 iterations.}

To investigate the performance of the methods under varying values of the design parameter $k$, three different values are specified for each method. Following the suggestions in \cite{zantek2006design}, we set for SSC the values $k_{SSC}=0.25$, 0.375, or 0.5, while for PRC we set the corresponding values $k_{PRC}=0.5$, 0.75, or 1. This means that we have $4 \times 10 \times 3 = 120$ scenarios in total for each method. Regarding the prior for PRC, we conduct a small sensitivity analysis by setting two priors to evaluate the performance of PRC under the absence or the presence of prior information. Therefore, we have two processes for PRC: 
\begin{itemize}
	\item PRC$_n$ for the non-informative reference prior  {$NIG(0,0,-1/2,0)$}
	\item PRC$_i$ for the weakly informative prior $NIG(0,4,2,1.5)$. 
\end{itemize}
 {Regarding the weakly informative prior, it is positioned at the correct values of the parameters and assigns equal information weight as four observations from the IC distribution. Specifically, if we assume four imaginary values $y_i$, where $i=1,...,4$, from the standard Normal and denoting their sample mean as $\bar{y}$, then from properties of Normal distribution we have $E \left( \bar{y}\right) =0$ and $E\left( \sum_{i=1}^4 \left( y_i - \bar{y} \right)^2 \right) = 3$, i.e., the expected sample mean is zero and the expected sum of squared deviations equal to three. Thus, we set $\mu_0=0$, $\lambda=4$, $\alpha=4/2$ and $\beta=3/2$ to illustrate the effect of the available prior information on the detection performance.} Note that this is the prior setting used in the simulation study in \cite{bourazas2023predictive}. The results are presented graphically in Figure \ref{fig:sim}, and they are tabulated in  {Table \ref{tab:sim}} in Appendix.

First of all, we observe that the larger the jump, the more rapidly all methods detect it. Additionally, both methods enhance their detection performance when the change point $\tau$ occurs later, i.e., when more IC data points are available to estimate the unknown parameters. For jumps of size 1, 1.5, and 2, the performance seems to stabilize beyond a certain point, implying that the methods do not require many data points to achieve their ``best'' performance for jumps of these magnitudes. However, this is not the case for the smaller jump of 0.5, where (for various values of  $k$) it takes time for $CED(\tau)$ to stabilize (or has not stabilized even with 100 IC data points). This indicates that a larger number of IC data points is needed for the efficient detection of small shifts. Notably, for $\delta=0.5$ when $\tau=11$, the detection ability appears to be very limited, as, in most cases, the $CED$ values are greater than 300, i.e., close to the nominal value of $ARL_0=370$. Comparing the methods, we notice the positive impact of the prior in decreasing CED for PRC$_i$, particularly when $\tau$ occurs early where only a limited number of data points are available. On the other hand, the performance of PRC$_i$ converges with PRC$_n$ and SSC when $\tau$ occurs later, and the volume of IC data increases, washing out the impact of the prior. Comparing PRC$_n$ and SSC, we observe that, generally, they demonstrate similar performances with small differences in CED. SSC has a slight advantage for small  {values of $k$s and $\tau$ occurs early in the process}. However, in those cases, the CED values are exceptionally high for both methods, rendering them ineffective for settings like short runs. On the other hand, PRC$_n$ shows slightly better performance for larger values of $k$s.

\begin{figure}[htbp]
	\centering
	\includegraphics[width=12cm]{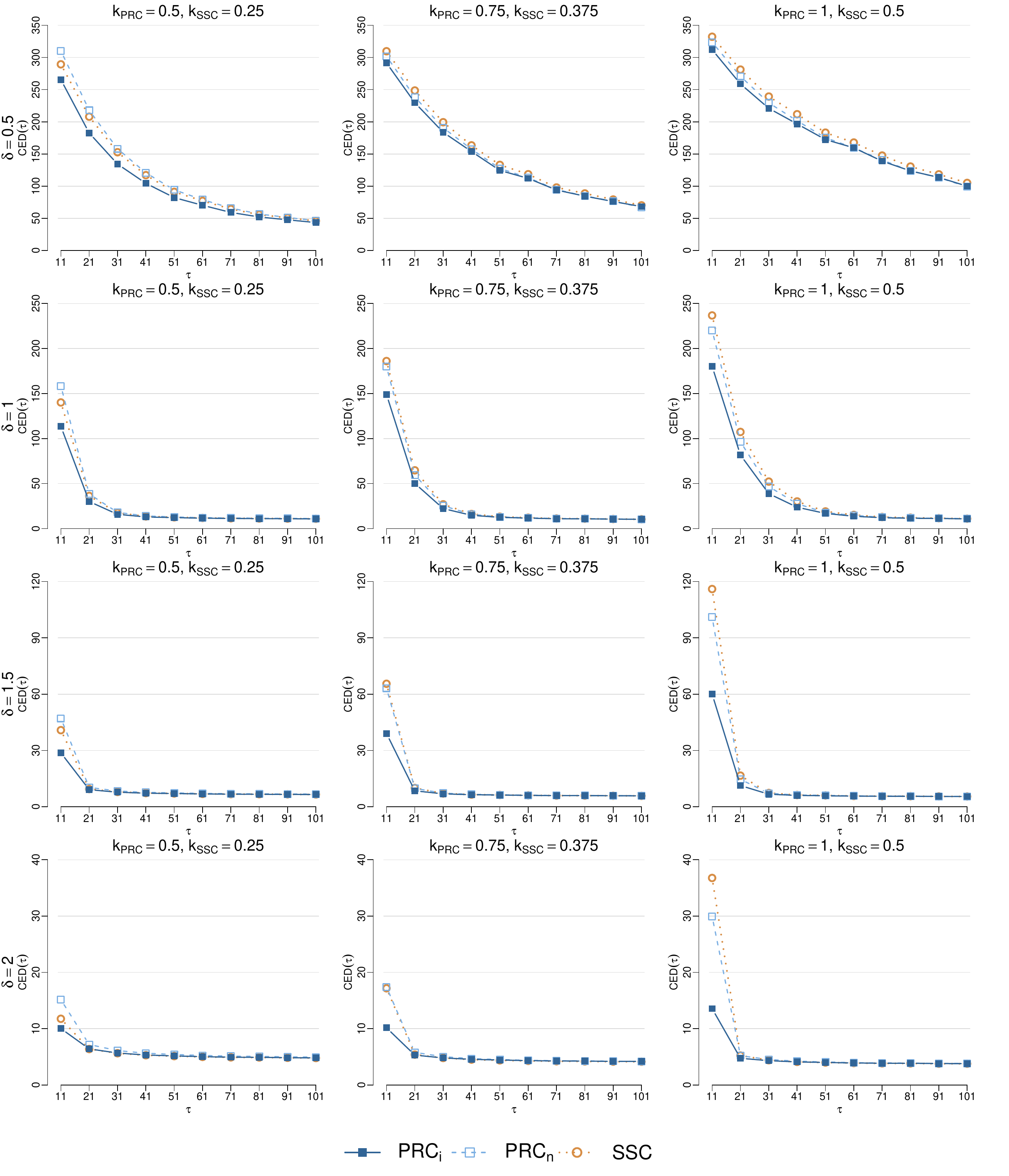}
	\caption{The conditional expected delay $CED(\tau)$ at locations $\tau = 11,\;21,\;...,\;101$, for shifts size of $\delta = 0.5$, $1$, $1.5$, or $2$ of SSC, and PRC, under the non-informative reference (PRC$_n$) or a moderately informative (PRC$_i$) prior. The design parameters are set $k_{PRC} = 0.5$, $0.75$, or $1$,  and $k_{SSC} = k_{PRC}/2$.}
	\label{fig:sim}
\end{figure}

\section{Conclusions} \label{sec:4}

In this study, the focus was on online testing for a change point in the mean of univariate Normal data without the presence of Phase I. Specifically, we evaluated and compared the CED performance for two parametric self-starting methods, the frequentist SSC, and the Bayesian PRC, under various magnitudes of shifts for the mean, and for a wide range of different change point positions. For PRC, we also conducted a small prior sensitivity analysis, assuming two priors: the non-informative reference prior and a weakly informative prior.

As expected, both methods detect large changes more quickly compared to small changes, and improve their performance as the IC horizon increases, or in other words, when $\tau$ appears at later stages. Moreover, large values of the design parameters $k$ enhance performance for significant changes but simultaneously increase the risk of absorbing smaller changes. Comparing the methods, we observe the beneficial effect of the prior in reducing CED, especially at the beginning of the process when only a few data points are available. Regarding the comparison of the PRC with the non-informative prior (PRC$_n$) against SSC, we see that they exhibit similar performances with very small differences in CED overall. SSC has a slight advantage by setting a small value for the design parameter $k$, while PRC performs better for larger values of $k$.


\newpage

\section*{Appendix}

\begin{table}[h]
	\tabcolsep9pt
	\begin{center}
		\resizebox{1\textwidth}{!}{
			\begin{tabular}{ccccccccccc} 
				&& \multicolumn{3}{c}{\large $k_{PRC}=0.5$, $k_{SSC}=0.25$} & \multicolumn{3}{c}{\large $k_{PRC}=0.75$, $k_{SSC}=0.375$} & \multicolumn{3}{c}{\large $k_{PRC}=1$, $k_{SSC}=0.5$} \\[4pt]
				$\delta$ &$\tau$ & $PRC_i$ & $PRC_n$ & $SSC$ & $PRC_i$ & $PRC_n$ & $SSC$ & $PRC_i$ & $PRC_n$ & $SSC$ \\ 
				\\[-1.8ex]\hline 
				\hline \\[-1.8ex] 
				\multirow{10}{*}{ {{\large $0.5$}} } & $11$ & $265.557$ & $310.294$ & $289.331$ & $291.529$ & $301.762$ & $309.652$ & $312.098$ & $323.628$ & $332.204$ \\ 
				& $21$ & $182.606$ & $217.874$ & $207.950$ & $229.683$ & $238.441$ & $248.640$ & $258.898$ & $270.766$ & $281.146$ \\ 
				& $31$ & $134.312$ & $158.108$ & $152.737$ & $183.739$ & $190.724$ & $199.355$ & $220.929$ & $229.446$ & $239.264$ \\ 
				& $41$ & $104.387$ & $120.656$ & $117.141$ & $153.817$ & $157.457$ & $163.413$ & $196.752$ & $201.528$ & $211.785$ \\ 
				& $51$ & $82.235$ & $94.272$ & $91.157$ & $124.477$ & $127.504$ & $133.084$ & $172.396$ & $175.305$ & $183.147$ \\ 
				& $61$ & $70.252$ & $79.018$ & $77.140$ & $112.104$ & $113.172$ & $118.558$ & $159.620$ & $159.514$ & $167.806$ \\ 
				& $71$ & $59.161$ & $65.934$ & $64.483$ & $94.000$ & $94.591$ & $97.963$ & $138.930$ & $140.639$ & $147.576$ \\ 
				& $81$ & $52.126$ & $56.685$ & $55.674$ & $84.490$ & $84.767$ & $88.688$ & $123.703$ & $123.719$ & $130.485$ \\ 
				& $91$ & $47.692$ & $51.401$ & $50.791$ & $76.172$ & $76.633$ & $79.237$ & $113.584$ & $113.323$ & $118.368$ \\ 
				& $101$ & $43.430$ & $46.278$ & $45.934$ & $68.223$ & $67.173$ & $70.161$ & $99.971$ & $99.039$ & $105.028$ \\ 
				\hline \\[-1.8ex]
				\multirow{10}{*}{  {{\large $1$}} } &  $11$ & $113.563$ & $158.367$ & $140.094$ & $148.950$ & $179.849$ & $186.042$ & $180.157$ & $219.954$ & $236.668$ \\ 
				& $21$ & $30.118$ & $38.499$ & $36.310$ & $50.252$ & $59.526$ & $64.747$ & $81.708$ & $96.471$ & $107.350$ \\ 
				& $31$ & $15.902$ & $18.064$ & $17.547$ & $22.315$ & $25.397$ & $27.012$ & $38.858$ & $46.752$ & $52.245$ \\ 
				& $41$ & $13.015$ & $14.119$ & $13.647$ & $14.825$ & $15.886$ & $16.256$ & $23.787$ & $27.662$ & $30.198$ \\ 
				& $51$ & $12.251$ & $12.968$ & $12.652$ & $12.543$ & $13.039$ & $13.258$ & $16.910$ & $17.771$ & $19.111$ \\ 
				& $61$ & $11.634$ & $12.213$ & $11.910$ & $11.700$ & $12.058$ & $12.189$ & $13.787$ & $14.687$ & $15.241$ \\ 
				& $71$ & $11.342$ & $11.829$ & $11.557$ & $11.011$ & $11.201$ & $11.259$ & $12.131$ & $12.694$ & $12.969$ \\ 
				& $81$ & $11.115$ & $11.566$ & $11.317$ & $10.904$ & $11.046$ & $11.078$ & $11.564$ & $11.808$ & $12.159$ \\ 
				& $91$ & $11.039$ & $11.411$ & $11.213$ & $10.604$ & $10.677$ & $10.690$ & $11.338$ & $11.497$ & $11.602$ \\ 
				& $101$ & $10.776$ & $11.145$ & $10.937$ & $10.335$ & $10.396$ & $10.415$ & $10.856$ & $11.054$ & $11.090$ \\ 
				\hline \\[-1.8ex] 
				\multirow{10}{*}{  {{\large $1.5$}} } &  $11$ & $28.762$ & $47.111$ & $40.836$ & $39.077$ & $63.196$ & $65.549$ & $60.090$ & $101.115$ & $116.028$ \\ 
				& $21$ & $9.124$ & $10.325$ & $9.463$ & $8.508$ & $9.787$ & $10.018$ & $11.327$ & $14.633$ & $16.613$ \\ 
				& $31$ & $7.783$ & $8.418$ & $7.934$ & $6.907$ & $7.280$ & $7.131$ & $6.705$ & $7.205$ & $7.358$ \\ 
				& $41$ & $7.250$ & $7.697$ & $7.341$ & $6.437$ & $6.632$ & $6.517$ & $6.070$ & $6.308$ & $6.292$ \\ 
				& $51$ & $7.062$ & $7.408$ & $7.120$ & $6.264$ & $6.396$ & $6.302$ & $5.880$ & $6.081$ & $6.058$ \\ 
				& $61$ & $6.847$ & $7.139$ & $6.894$ & $6.091$ & $6.177$ & $6.114$ & $5.699$ & $5.792$ & $5.774$ \\ 
				& $71$ & $6.727$ & $6.975$ & $6.765$ & $5.959$ & $6.023$ & $5.966$ & $5.619$ & $5.689$ & $5.660$ \\ 
				& $81$ & $6.678$ & $6.911$ & $6.723$ & $5.964$ & $6.012$ & $5.963$ & $5.612$ & $5.662$ & $5.652$ \\ 
				& $91$ & $6.612$ & $6.825$ & $6.656$ & $5.906$ & $5.931$ & $5.884$ & $5.517$ & $5.555$ & $5.544$ \\ 
				& $101$ & $6.533$ & $6.727$ & $6.564$ & $5.853$ & $5.868$ & $5.829$ & $5.475$ & $5.525$ & $5.507$ \\ 
				\hline \\[-1.8ex]
				\multirow{10}{*}{ {{\large $2$}} } & $11$ & $10.018$ & $15.146$ & $11.752$ & $10.152$ & $17.401$ & $17.170$ & $13.562$ & $29.935$ & $36.761$ \\ 
				& $21$ & $6.405$ & $7.179$ & $6.380$ & $5.302$ & $5.781$ & $5.390$ & $4.739$ & $5.189$ & $5.222$ \\ 
				& $31$ & $5.670$ & $6.099$ & $5.631$ & $4.799$ & $5.020$ & $4.794$ & $4.309$ & $4.523$ & $4.392$ \\ 
				& $41$ & $5.320$ & $5.636$ & $5.273$ & $4.554$ & $4.679$ & $4.502$ & $4.095$ & $4.220$ & $4.101$ \\ 
				& $51$ & $5.158$ & $5.409$ & $5.120$ & $4.425$ & $4.514$ & $4.371$ & $3.993$ & $4.079$ & $3.989$ \\ 
				& $61$ & $5.027$ & $5.233$ & $4.986$ & $4.320$ & $4.373$ & $4.274$ & $3.904$ & $3.959$ & $3.892$ \\ 
				& $71$ & $4.929$ & $5.118$ & $4.902$ & $4.261$ & $4.301$ & $4.215$ & $3.854$ & $3.901$ & $3.834$ \\ 
				& $81$ & $4.906$ & $5.071$ & $4.881$ & $4.234$ & $4.262$ & $4.182$ & $3.849$ & $3.877$ & $3.826$ \\ 
				& $91$ & $4.841$ & $4.996$ & $4.824$ & $4.191$ & $4.205$ & $4.138$ & $3.801$ & $3.818$ & $3.771$ \\ 
				& $101$ & $4.806$ & $4.945$ & $4.785$ & $4.176$ & $4.180$ & $4.126$ & $3.791$ & $3.811$ & $3.770$ \\ 
				\hline \\[-1.8ex]		
			\end{tabular} \label{tab:sim}} 
   		\caption{The conditional expected delay ${CED(\tau)}$ at locations ${\tau = 11,\;21,\;...,\;101}$, for shifts size of $\delta = 0.5$, $1$, $1.5$, or ${2}$ of SSC, and PRC, under the non-informative reference (PRC$_n$) or a moderately informative (PRC$_i$) prior. The design parameters are set ${k_{PRC} = 0.5}$, ${0.75}$, or ${1}$,  and ${k_{SSC}} = {k_{PRC}/2}$. }
	\end{center}
\end{table}

\end{document}